**Where's the Bus? Examining Factors Correlated with High and Low Public Transit On-Time Performance.**

**Snehanshu Banerjee**
Senior Transportation Engineer
Omni Strategy, Baltimore, MD 21201
Email: snban1@morgan.edu

**Mohammadreza Jabehdari**
Department of Industrial & Systems Engineering
Morgan State University, Baltimore, MD 21251
Email: mojab1@morgan.edu

**Ross Turlington**
Manager, External Reporting and Policy Control
Maryland Transit Administration
Office of Performance Management
Baltimore, MD 21202
Email: rturlington@mta.maryland.gov

**Cole Greene**
Chief Performance Officer
Maryland Transit Administration
Office of Performance Management
Baltimore, MD 21202
Email: cgreene@mta.maryland.gov

## ABSTRACT

An important aspect of public bus transit is its reliability of being on-time, which has a major impact on bus ridership. Currently, there is no unified national standard to measure bus on-time performance in the United States. This paper proposes a novel approach to improve the on-time performance (OTP) of buses at transit agencies throughout the country, based on factors unique to each agency. In this study, information on 18 public transit agencies data were obtained from the National Transit Database and were verified by the authors after contacting the respective agencies. A correlation analysis was conducted to evaluate the influence of different factors on bus OTP. The results show that a higher "congestion ranking" of the cities i.e. lesser traffic has a positive correlation with OTP while a lower number of passengers per hour has a negative correlation with OTP, both at a 95% confidence interval. "Total miles per active vehicle" and "average speed" of the bus were found to have a positive correlation with OTP at a 90% confidence interval. This potentially suggests that, if the average speed of the bus is higher i.e. if the bus operates at its scheduled speed, total miles per active vehicle will increase, which will in turn lead to a better on-time performance. A flow diagram interpreting the results of the correlation analysis was shown, where the main bus service process was considered and the predecessors and successors which could potentially affect the final OTP, were determined.

**Keywords:** On-Time Performance, Public Transit Bus, Transit Agencies, National Transit Database

## INTRODUCTION

The increasingly congested urban environments of today demand that transportation planners primarily focus their efforts on improving public transit modalities as an alternative to single occupancy vehicles, thereby decrease congestion [1]. Modal public transit options provide a more sustainable way of transporting people in urban areas due to their ability to move large groups of passengers at consistent intervals. Additionally, public transit reduces road congestion, air pollution, and energy consumption [2, 3]. Realizing this, several State Departments of Transportation (DOTs) and local transportation agencies have begun to direct their efforts and financial resources towards improving public transportation services with the goal of attracting more riders and relieving congestion [1, 3]. Since 1990, the funding allocated for improving public transit has increased considerably [4], from $26.1 billion in 1997 [1], to $70 billion in 2017 [5].

Although more resources and emphasis have been allocated to developing public transit, national ridership has experienced declines in recent years [6]. Among the most important factors impacting public transit ridership is reliability [7]. Reliability is defined as "the degree to which the actual service reflects the scheduled service" [8]. From the perspective of the passenger, a reliable public transit mode is the one that best adheres to schedules, provides consistent travel times, and offers short waiting and on-board vehicle travel times [8, 9]. To best assess these measures, several parameters are used to benchmark the reliability of public transit modes or routes, with on-time performance being the most common [1]. According to Fan, Guthrie and Levinson [10], the on-time performance of public transit is the most important factor that impacts transportation mode choice in the developed world, outweighing even the cost associated with that specific mode. Consequently, public transportation agencies and state DOTs are continuously trying to improve the on-time performance of their public transportation services to attract more passengers [3, 11]. These agencies have established key performance indicators (KPIs) and implemented several practices to measure the on-time performance of their different public transit modalities. Despite the importance and ubiquity of on time performance as a KPI, there is no unified national standard for this measure [12]. Each transportation agency applies its own definition of "on-time" performance to the varying public transportation modes it operates. Complicating matters further, the same agency, in some cases, applies different on time performance definitions, and goals to each of their public transit modes. In this first of a kind study, the authors collected bus on-time performance data and other associated factors from the major transit agencies in the United States, all of which could possibly affect on-time performance and evaluated them for any kind of correlation. Potentially, this could help create standardized transit performance measures, from which transit agencies can use to benchmark one another and make operational improvements to better serve transit passengers.

## Background

In the following section, the KPIs used by different transportation agencies across the United States to measure their transit mode on-time performances will be presented, including their existing performances and measurement criterions.

### *On-time performance definition of public transit*

A majority of the transportation agencies studied for this paper use a standard method to capture on time performance information through the use of Automatic Vehicle Locator systems (AVL) which collects actual transit bus arrival and departure time data [13]. AVLs are "computer-based vehicle tracking systems that measure the real-time position of each vehicle and relay this information back to a central location" [13]. These systems use either GPS, ground-based radio, signpost and odometer interpolation, or dead reckoning to track the vehicles, with GPS being the most effective tracking system [13].

Among public transit buses, the most-widely used definition for on-time performance considers the bus to be "on-time" if it arrives at a scheduled stop no earlier than one minute before, and no later than five minutes after the times identified on the official published bus schedules. This definition is used by public transportation agencies in the cities of Los Angeles, CA [14], Chicago, IL [15], Seattle, WA [16], Denver, CO [17], Portland, OR [18], Minneapolis/Saint Paul, MN [19], Miami-Dade, FL [20], Pittsburgh, PA [21],

Santa Clara, CA [22], and New Orleans, LA [23]. A similar time interval, but with different parameters is applied by the Southeastern Pennsylvania Transportation Authority (SEPTA) in southeastern Pennsylvania, Capital Metro in Austin, TX, and Niagara Frontier Transportation Authority (NFTA) in Buffalo, NY. According to SEPTA [24] and Capital Metro [25], a bus is considered "on-time" if it arrives at a scheduled stop no earlier than one minute before, and no later than six minutes while NFTA [26] considers a bus to be on time if it is less than two minutes early and less than four minutes late, i.e. a six minute interval like SEPTA and Capital Metro, after the times identified on the official published bus schedules.

Several other transportation agencies have adopted a stricter version of the above definitions when assessing on-time performance of the public transit buses. Two examples include transportation agencies in the cities of San Francisco, CA and the State of Missouri. According to San Francisco Municipal Transportation Agency (SFMTA) [27] and the Missouri DOT [28], buses are considered to be "on-time" if it arrives at a scheduled stop no earlier than one minute before, and no later than four minutes after the times identified on the official published bus schedules. Similarly, in Phoenix, AZ, the state of Utah, and Charlotte, NC, the public transportation authority's use a five-minute interval to define the bus "on-time" performance; albeit, the interval is only applicable to the bus arriving late at the stop with no early arrival allowed [22, 29, 30].

By contrast, among the surveyed jurisdictions, Washington Metropolitan Area Transit Authority (WMATA) and Maryland Transit Administration (MTA) have the most relaxed definitions of "on-time" for measuring transit buses performance. WMATA [31] and MTA [32] consider a bus to be "on-time" if it arrives at a scheduled stop no earlier than two minutes before, and no later than seven minutes after the times identified on the official published bus schedules. This allows for a maximum window of nine minutes which a bus has, to arrive at a scheduled stop and still be considered "on-time".

The variance of "on-time" definitions are all related to the arrival of buses at a designated stop; however, a number of external factors may impact the bus arrival time; such as: weather conditions, pedestrian commotion and police activity [33]. Several transportation authorities measure departure time from a stop, as their "on-time" performance. For example, the state of New Jersey [34], Houston, TX [35], Atlanta, GA [36], and Las Vegas, NV [37], all consider the bus to be "on-time" if it departs a designated stop anytime between zero and six minutes after the published schedule. Similarly, Boston, MA uses departure time to define "on-time" performance, but with a tighter interval of only three minutes later than the scheduled time for high frequency bus routes; while it uses a more relaxed interval of one minute before and six minutes after for infrequent bus routes [38].

Two other cities use measures, other than departure and arrival times to measure transit bus on-time performance. The transportation agency of Honolulu, HI uses headways to measure the on-time performance of its transit buses. Headways measure the integrity of time between two buses operating the same route. Accordingly, a Honolulu transit bus is considered to be "on-time" if it arrives at a stop within a +/- three minutes of scheduled headway of a bus operating the same route [39]. Perhaps, the most comprehensive definitions for "on-time" performance comes from New York City (NYC). NYC measures its on-time performance in three distinct ways [40]:

1- *Customer Journey Time Performance:* This measure represents the percentage of passengers who complete their journey within 5 minutes of the scheduled time. This KPI is measured using bus and subway card swipes coupled with vehicle GPS data.
2- *Additional Travel Time*: This is the average additional time that a passenger would spend onboard the bus beyond its scheduled arrival time.
3- *Additional Bus Stop Time:* Measures the average added time passengers wait at a bus stop.

Due to the lack of unified standards for "on-time" performance, different public transportation agencies use unique definitions to measure this KPI; summarized below in Table 1;

**TABLE 1 Public Buses "On-Time" Performance Definitions**

| Measure | Arrival/Departure | Time Interval | State/City |
|---|---|---|---|
| OTP | Arrival | One minute early to five minutes late | Los Angeles, CA; Chicago, IL; Seattle, WA; Denver, CO; Portland, OR; Minneapolis/Saint Paul, MN; Miami-Dade, FL; Pittsburgh, PA; Santa Clara, CA; New Orleans, LA. |
| | | Two minutes early to four minutes late | Buffalo, NY |
| | | Zero to six minutes late | Southeastern Pennsylvania; Austin, TX. |
| | Arrival | One minute early to four minutes late | San Francisco, CA, State of Missouri |
| | | Zero to five minutes late | Phoenix, AZ; State of Utah, Charlotte, NC. |
| | | Two minutes early to seven minutes late | Washington DC |
| OTP | Departure | Zero to six minutes late | New Jersey; Houston, TX; Atlanta, GA; Las Vegas, NV. |
| | | Zero to three minutes late for frequent routes | Boston, MA |
| | | One minute early to six minutes late for infrequent routes | |
| Headway | Arrival | +/- three minutes of scheduled headway | Honolulu, HI |
| Customer Journey Time Performance | Arrival | Zero to five minutes late | New York City |

## METHODOLOGY

Several agencies publish the actual on-time performance data for its different public transportation modes. In this section, a comparison of these actual performances for the public transportation buses, using the latest available data from each transportation agency, will be conducted. Regarding the on-time performance for public buses, Table 2 lists the year in which the on-time performance of the public bus service of different agencies across the US was measured as well as the sources of this data; while, Figure 1 shows a comparison of the latest available data from the different public transportation agencies.

**TABLE 2 Public Buses "On-Time" Performance Measurement Dates and Sources**

| Agency | Date of Measurement | Source |
|---|---|---|
| Missouri | 2015 | [28] |
| Denver | 3rd quarter 2016 | [17] |
| Utah | 2017 | [30] |
| New Jersey | 2013 | [34] |
| Phoenix | 2018 | [22] |
| Portland | March 2019 | [18] |
| Las Vegas | February 2018 | [37] |
| Santa Clara | 2nd quarter 2019 | [22] |
| Austin | February 2019 | [25] |
| Buffalo | 2018 | [26] |
| Miami-Dade | January 2018 | [20] |
| Seattle | October 2018 | [16] |
| Atlanta | February 2019 | [36] |
| Houston | YTD 2019 (February) | [35] |
| New Orleans | 2018 | [23] |
| New York City | March 2019 | [40] |
| Honolulu | 8 months till February 2019 | [39] |
| Maryland | January 2019 | [32] |
| Pittsburgh | 2017 | [21] |
| Boston | April 2019 | [38] |
| San Francisco | April 2019 | [27] |

The table above identifies transit properties by their corresponding urbanized zone area and includes the sources and dates for the on-time performance figures cited herein.

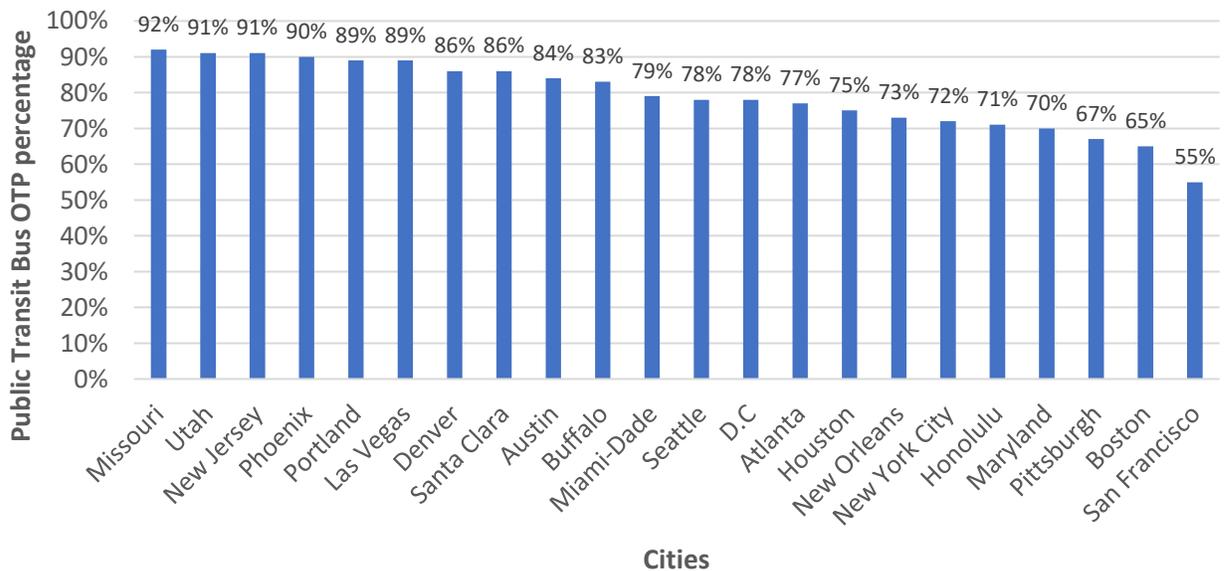

**Figure 1 Transit agency bus on-time performance**

Figure 1 illustrates bus on-time performance percentages for several transportation agencies in the respective cities. The on-time performances vary from as high as 92% in the state of Missouri to as low as 55% in San Francisco, CA. The range of variations is in part due to the adopted definitions for OTP. Ten different cities achieved more than 80% on-time performance, which is a generally acceptable level for public transit bus service. Furthermore, apart from San Francisco, the worst performers were Pittsburgh and Boston with 67% and 65% on-time performance, respectively. Moreover, of the agencies that use the most common on-time definition, New Orleans had the worst performance with only 73% on-time performance for public transit buses. The time interval used by different transit agencies as shown in Table 1, would imply that wider time intervals would lead to better OTP, however analysis is needed to prove the hypothesis that there is a correlation between these factors.

**National Transit Database**

The Federal Transit Administration (FTA) publishes the National Transit Database (NTD) to record the financial, operating and asset conditions of all transit systems in the United States, which is publicly available. Some of the terms used in the NTD and mentioned in this study are explained below:

1. *Agency VOMS*: "The number of revenue vehicles operated across the whole agency to meet the annual maximum service requirement. This is the revenue vehicle count during the peak season of the year; on the week and day that maximum service is provided. Vehicles operated in maximum service (VOMS) exclude atypical days and one-time special events".

2. *Active Vehicles*: "The vehicles available to operate in revenue service at the end of the fiscal year, including:
   • Spares
   • Vehicles temporarily out of service for routine maintenance and minor repairs
   • Operational vehicles".

3. *Total Miles on Active Vehicles during the Period*: "The total miles accumulated during the period on all active vehicles, based on the end of period inventory".

4. *Trip Operating Expenses*: "The expenses associated with the operation of the transit agency, and classified by function or activity, and the goods and services purchased".

5. *Major Failures*: "A failure of some mechanical element of the revenue vehicle that prevents the vehicle from completing a scheduled revenue trip or from starting the next scheduled revenue trip because actual movement is limited or because of safety concerns".

6. *Type of Service*: "Describes how public transportation services are provided by the transit agency: directly operated (DO) or purchased transportation (PT) services".

7. *Bus (MB)*: "A transit mode comprised of rubber-tired passenger vehicles operating on fixed routes and schedules over roadways. Vehicles are powered by: Diesel, Gasoline, Battery, Alternative fuel engines contained within the vehicle".

8. *Unlinked Passenger Trips*: "The number of passengers who boarded public transportation vehicles. Passengers are counted each time they board a vehicle no matter how many vehicles they use to travel from their origin to their destination".

9. *Vehicle Revenue Hours*: "Total number of hours that vehicles/passenger cars traveled while in revenue service during the report year. Includes both typical and atypical service. Excludes deadhead".

10. *Vehicle Revenue Miles*: "The miles that vehicles (or passenger cars, for rail service) travel while in revenue service. Vehicle revenue miles exclude deadhead, operator training, maintenance testing, and school bus and charter services".

11. *Passengers per Hour*: "The average number of passengers to board a vehicle/passenger car in one hour of service".

12. *Traffic congestion*: "A condition on road networks that occurs as use increases, and is characterized by slower speeds, longer trip times, and increased vehicular queueing".

The authors compiled a list of metrics from NTD, corroborated the agency bus OTP and interval times found in the literature with individual agency follow ups and used the congestion ranking of cities (higher the ranking, lesser the congestion) from the INRIX report [41]. This compiled list is shown in Table 2. Boston has the worst congestion, whereas Las Vegas has the least congestion compared to the cities in Table 2.

**TABLE 2 Public transit bus related information from NTD database and INRIX report**

| Transit Agencies | Total miles /Active vehicle | Trip operating expenses per unlinked passenger trip | Avg Speed | Congestion ranking | Vehicles operated in annual max service | Passengers /hr | Unlinked passenger trips/ Vehicles operated in annual max service | Total miles/ Major failure | Failure/ fleet | Inter val | OTP |
|---|---|---|---|---|---|---|---|---|---|---|---|
| Honolulu | 39,232 | $2.98 | 12.9 | 18 | 452 | 47.39 | 144,416 | 9243 | 4.17 | 3 | 71% |
| Boston | 20,520 | $4.12 | 9.8 | 1 | 767 | 47.71 | 498,926 | 14001 | 1.38 | 3 | 65% |
| Phoenix | 43,196 | $4.09 | 11.1 | 22 | 451 | 23.05 | 86,471 | 12258 | 3.88 | 5 | 90% |
| Portland | 30,086 | $4.39 | 12.5 | 10 | 29 | 22.53 | 236,907 | 19630 | 1.23 | 6 | 89% |
| Las Vegas | 44,684 | $2.40 | 11.2 | 31 | 278 | 49.82 | 227,203 | 24604 | 1.69 | 6 | 89% |
| New York City | 25,115 | $3.65 | 7.0 | 4 | 3,257 | 56.66 | 1,056,384 | 11230 | 2.30 | 5 | 72% |
| Denver | 39,387 | $5.22 | 13.2 | 19 | 867 | 44.81 | 75,278 | 68609 | 0.61 | 6 | 86% |
| Austin | 40,242 | $5.34 | 11.8 | 14 | 314 | 22.19 | 86,933 | 5489 | 7.05 | 6 | 84% |
| Buffalo | 31,949 | $4.62 | 11.0 | 29 | 271 | 27.89 | 79,714 | 13433 | 2.22 | 6 | 83% |
| Miami-Dade | 39,226 | $6.26 | 11.5 | 12 | 709 | 23.52 | 81,807 | 19672 | 2.13 | 6 | 79% |
| Seattle | 32,918 | $5.12 | 11.0 | 6 | 986 | 33.99 | 129,771 | 5093 | 6.29 | 6 | 78% |
| Atlanta | 53,642 | $3.63 | 12.4 | 11 | 466 | 27.18 | 123,305 | 4252 | 12.82 | 6 | 77% |
| Houston | 34,400 | $5.51 | 12.1 | 13 | 591 | 20.75 | 82,906 | 73167 | 0.33 | 6 | 75% |
| New Orleans | 45,279 | $5.83 | 12.2 | 26 | 90 | 22.68 | 116,691 | 10037 | 4.47 | 6 | 73% |
| Pittsburgh | 35,504 | $5.64 | 13.1 | 7 | 614 | 33.17 | 102,981 | 15836 | 2.30 | 6 | 67% |
| San Francisco | 26,526 | $3.05 | 7.8 | 8 | 477 | 56.44 | 474,344 | 8756 | 2.90 | 5 | 55% |
| D.C | 30,176 | $5.13 | 10.1 | 2 | 1,290 | 31.18 | 273,291 | 11175 | 3.09 | 9 | 78% |
| Maryland | 31,006 | $4.34 | 11.4 | 15 | 604 | 40.09 | 115,785 | 5733 | 5.24 | 9 | 70% |

## RESULTS

A Pearson's correlation analysis was conducted to measure the strength of linear relationships between the different variables. As shown in Table 3, congestion ranking of the cities and passengers per hour were found to have statistically significant correlation with on-time performance at a 95% confidence interval.

**TABLE 3 Pearson's Correlation Analysis and Statistical Significance**

| | Total miles /Active vehicle | Trip operating expenses per unlinked passenger trip | Avg Speed | Congestion ranking | Vehicles operated in annual max service | Passengers/Hour | Unlinked passenger trips/ Vehicles operated in annual max service | Total miles/ Major failure | Interval |
|---|---|---|---|---|---|---|---|---|---|

| | | | | | | | | |
|---|---|---|---|---|---|---|---|---|
| Trip operating expenses per unlinked passenger trip | 0.031 | | | | | | | |
| *ρ value* | 0.902 | | | | | | | |
| Avg Speed | 0.607 | 0.361 | | | | | | |
| *ρ value* | 0.008* | 0.141 | | | | | | |
| Congestion ranking | 0.571 | -0.144 | 0.381 | | | | | |
| *ρ value* | 0.013* | 0.568 | 0.118 | | | | | |
| Vehicles operated in annual max service | -0.422 | -0.071 | -0.637 | -0.497 | | | | |
| *ρ value* | 0.081** | 0.78 | 0.004* | 0.036* | | | | |
| Passengers/Hour | -0.422 | -0.664 | -0.551 | -0.172 | 0.469 | | | |
| *ρ value* | 0.081** | 0.003* | 0.018* | 0.494 | 0.049* | | | |
| Unlinked passenger trips/Vehicles operated in annual max service | -0.583 | -0.409 | -0.836 | -0.468 | 0.795 | 0.655 | | |
| *ρ value* | 0.011* | 0.092** | 0* | 0.05* | 0* | 0.003** | | |
| Total miles/Major failure | 0.03 | 0.26 | 0.306 | 0.153 | -0.02 | -0.068 | -0.193 | |
| *ρ value* | 0.906 | 0.298 | 0.217 | 0.544 | 0.937 | 0.788 | 0.442 | |
| Interval | 0.079 | 0.357 | 0.094 | 0.006 | -0.004 | -0.32 | -0.274 | 0.008 |
| *ρ value* | 0.755 | 0.146 | 0.711 | 0.982 | 0.987 | 0.196 | 0.271 | 0.975 |
| OTP | 0.456 | 0.107 | 0.416 | 0.494 | -0.197 | -0.475 | -0.383 | 0.227 | 0.206 |
| *ρ value* | 0.057** | 0.674 | 0.086** | 0.037* | 0.433 | 0.046* | 0.116 | 0.364 | 0.413 |

* $\rho \leq 0.05$, ** $\rho \leq 0.1$

Passengers per hour has a negative correlation with OTP; which means that a lower the number of passengers per hour, increases the possibility of improving OTP as shown in Figure 2.

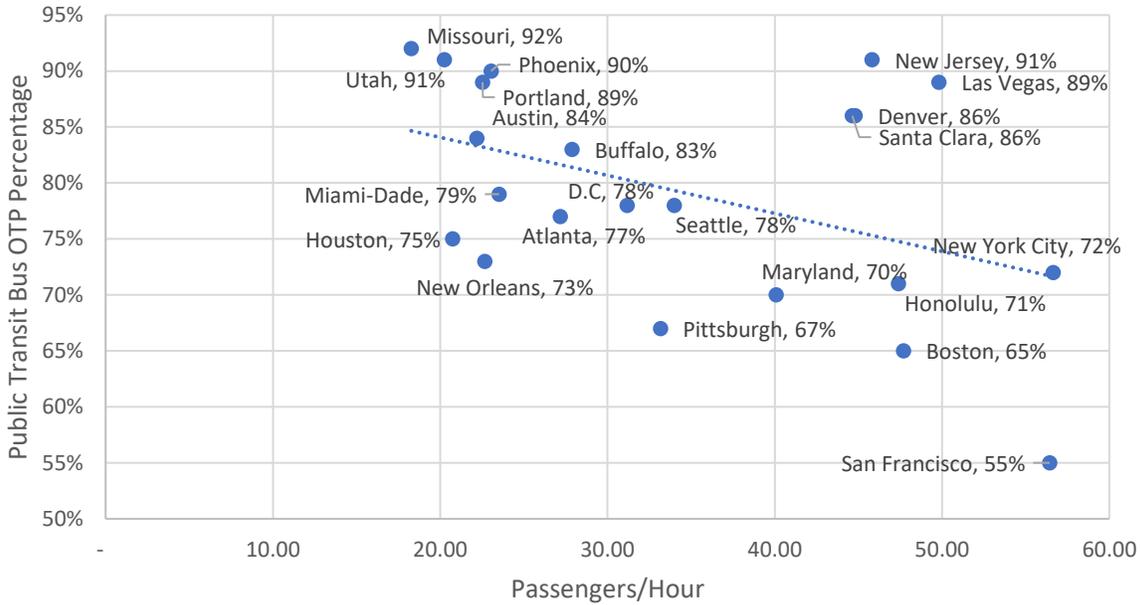

**Figure 2 Passengers per hour vs OTP**

This indirectly means that, if the active fleet size or bus availability is more, i.e. lesser bus failures would mean on schedule service, compared to a reduced bus avaiaibility, leading to cut trips and passengers queueing up for the next scheduled bus arrival, which impacts on-time performance.

Congestion has a positive correlation with OTP, which indicates that the higher the congestion ranking (on a worse to best scale) of the city, the better the OTP as shown in Figure 3.

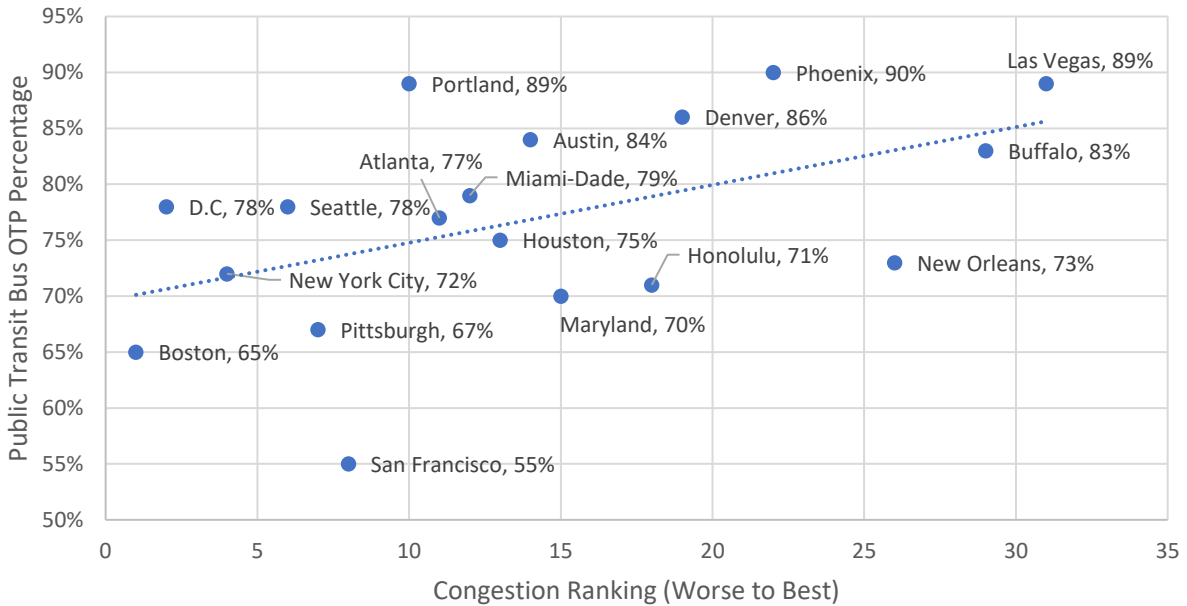

**Figure 3 Congestion ranking vs OTP**

Total miles per active vehicle and average speed also have a positive correlation with OTP at a 90% confidence interval and are shown in Figure 4 and Figure 5.

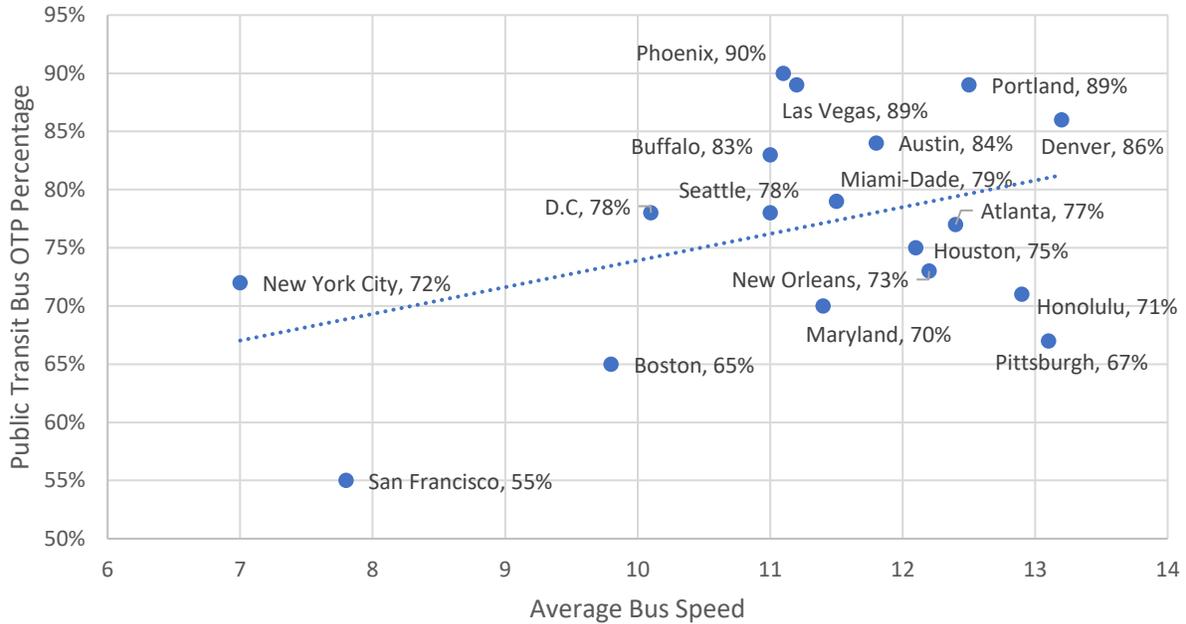

**Figure 4 Average bus speed vs OTP**

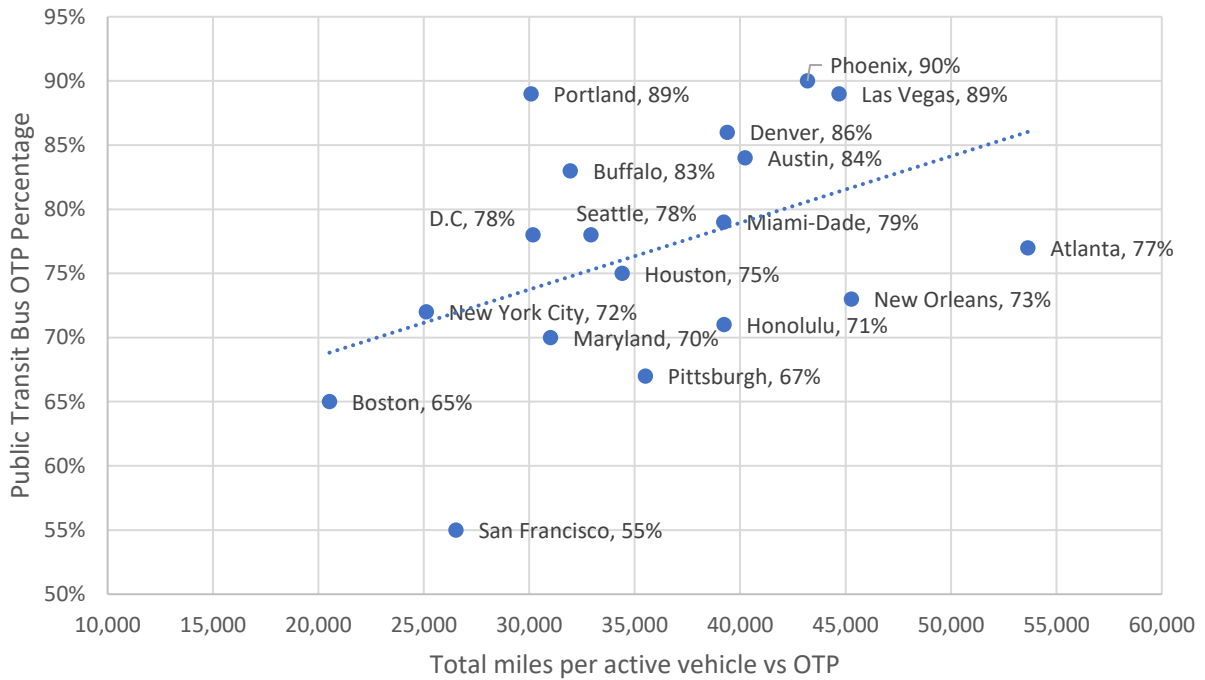

**Figure 5 Total miles per active vehicle vs OTP**

Other than OTP, the correlation analysis also shows variables that indirectly correlate to OTP, such as a higher average bus speed i.e. being at the scheduled bus speed, would lead to the estimated/scheduled miles per active vehicle.

**DISCUSSION**

Before improvement can be undertaken, the question of: "What are the factors involved in this issue and how much is their effect on having this OTP rate?" must first be addressed. A high-level conceptual model of bus transit service can clarify the lifecycle processes of interactions which directly or indirectly impact on-time performance.

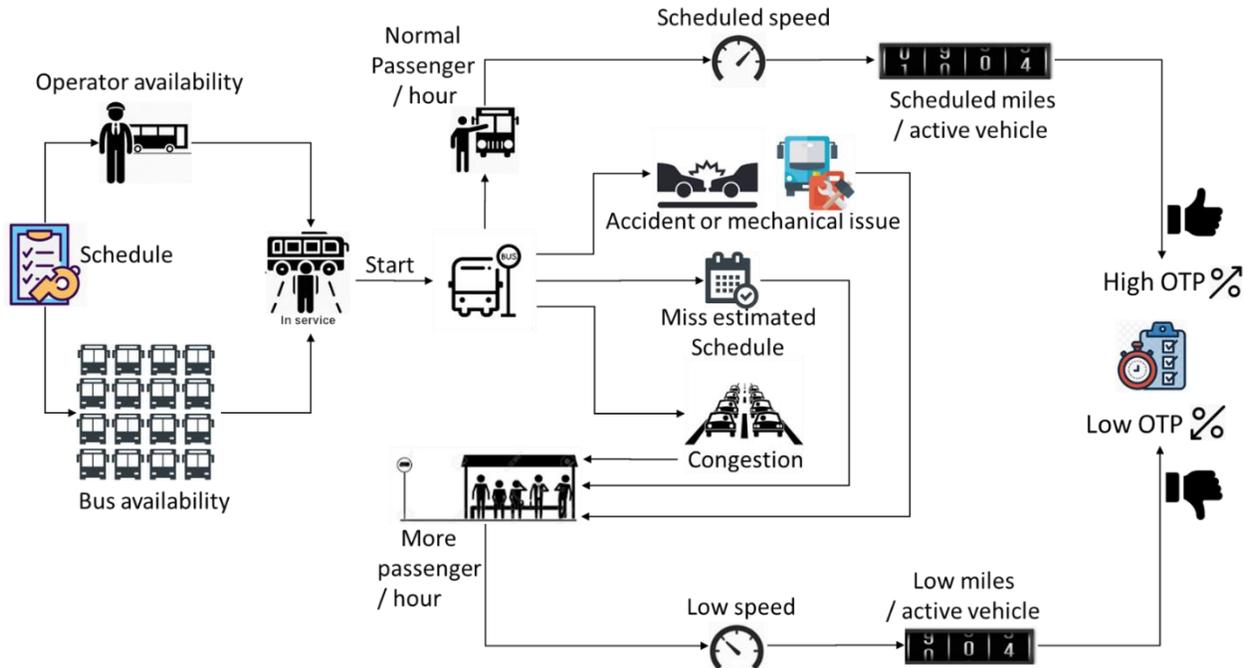

**Figure 6 Bus process flow diagram**

Figure 6 explains the correlation between a variety of factors and OTP. Some metrics are needed to measure the performance for each individual factor. Historical data can also be utilized to compare agencies against one another to measure the impacts of each factor. The information provided in Figure 6 looks to establish corollary patterns which help explain relationships between the impacting factors and OTP. Based on the correlation analysis, there are four primary factors affecting OTP. However, a multitude of additional variables may affect these four factors. For example, if there are not enough buses available, not only will OTP decrease but, it will also result in greater passenger wait times and the average speed would decrease.

In order to implement process improvements, transit agencies need to be able to identify best practices which have yielded positive results in other similar transportation environments and have benefitted the "Passenger per Hour" and "Average Speed" metrics. Figure 6 shows how these factors are involved in the whole process and what are the predecessors and successors tasks. One of the most important factors which can help to improve total miles per active vehicle, average speed, and passengers per hour is accurate schedules. Unscheduled bus allocations or corrective maintenance are the two most important influencing factors affecting "Bus Availability". There are several actions which can be taken to help improve the on-time performance of public transit buses [13]. Using real-time data available through the GPS tracking systems is an important technology to acquire and utilize. Additionally, training dispatchers to monitor the buses' arrival and departure times from specific stops and to adjust the published schedules to more accurately represented the behavior of buses. For example, Portland, OR makes quarterly adjustments to their bus schedules that are often behind to try constantly increase the accuracy of their published transit information [42]. Taking these steps will help improve the passenger´s perception of transit bus reliability and improve confidence in the overall transportation system.

Attempting to mitigate factors which increase variability in the on-time performance can help improve indicator significance [13]. Although some of these factors such as weather conditions cannot be controlled, others factors such as delay due to accidents or increased traffic can be managed efficiently using dedicated bus lanes – Seattle and Minneapolis/Saint Paul [43], and prioritizing traffic signals to improve schedule adherence [44, 45]. A median separated dedicated bus lane segregates transit buses from single occupancy vehicle traffic, preventing it from having to pull in and out of traffic to load passengers. These improvements will help increase the accuracy of the bus schedules and, in return, increase its reliability and passenger confidence. Transit signal priority facilitates the movement of transit vehicles through intersections by giving priority to these transit vehicles via increased green light times and traffic signal coordination; these improvements lead to improved OTP. Expanding rapid bus routes which contain fewer stops along a route helps increase ridership and improve reliability. In Richmond, VA, monthly ridership in 2018 increased by 21% compared to 2017 as a result of opening a new seven-mile rapid bus line, Richmond buses experienced greater reliability and better on-time performance compared to non-rapid transit routes [46].

Adopting fare payment policies which allow patrons to pay at each point of egress from a transit bus can also be an influential factor in mitigating dwell times and improving OTP [47]. By performing regularly scheduled maintenance to buses, providing operator training and implementing enhanced wayfinding and signage helps improve on-time performance of public transit buses [42]. Redesigning bus routes helps to increase bus speeds this results in greater schedule adherence accuracy. Eliminating branching and deviations, shortening the routes, and consolidating stops are additional ancillary considerations which help to increase OTP [48, 49]. Redesigning curbs can assist transit buses pulling in and out from their stops into traffic which will help buses adhere to their schedules improve travel speed [50].

Through analyzing data and systems from other agencies, transit agencies can determine patterns which prove relationships between various factors and OTP. To address improving OTP, a statistical correlation analysis has been tested, to determine the significance of each factors effect on OTP.

## CONCLUSIONS

The importance of public transportation in our modern urban cities is increasing every day. Public transportation provides a sustainable way of transporting people around urban environments, transit services have the capability of moving large numbers of passengers at high frequency intervals, with reduced impact to traffic and with greater environmental responsibility. In an attempt to increase public transit ridership, especially among public transit buses, transportation agencies across the country are trying to improve the reliability of these buses by enhancing their on-time performance.

This study provides an analysis from which transit agencies can learn which factors most significantly impact bus service reliability and efficiency. However, the lack of a standard and unified on-time performance definitions and measures across the country has resulted in convoluted normalization practices and increased the difficulty in performing OTP analysis. From as large as a nine-minutes interval in Washington DC to as small as the three-minutes in Boston, MA, there is a wide array of definitions of on-time performance as summarized in Table 1. By using the most applicable definition and setting the most realistic target for a particular city or state will depend on the unique characteristics of each urban area. This study demonstrates that most cities are able to achieve high on-time performance rates by adopting specific strategies. Individualized transit performance improvement plans can be generated based on a deeper understanding of the variables which influence fluctuations in transit service performance.

## ACKNOWLEDGMENTS

The authors would like to thank the Maryland Transit Administration, NJ Transit, Santa Clara Valley Transportation Authority, Massachusetts Bay Transportation Authority, Regional Transportation Commission of Southern Nevada, Houston Metro, Regional Transportation District-Denver, Metropolitan Atlanta Rapid Transit Authority, Tri-County Metropolitan Transportation District of Oregon and Utah Transit Authority, for their cooperation and explaining their bus on-time performance measures and other metrics in detail.

**AUTHOR CONTRIBUTIONS**
The authors confirm contribution to the paper as follows:
study conception and design: S. Banerjee, C. Greene;
data collection: S. Banerjee, M. Jabehdari;
analysis and interpretation of results: S. Banerjee, M Jabehdari, R. Turlington;
draft manuscript preparation: S. Banerjee, R. Turlington, M. Jabehdari, C. Greene.
All authors reviewed the results and approved the final version of the manuscript.

**REFERENCES**
1. Diab, E.I., M.G. Badami, and A.M. El-Geneidy, *Bus transit service reliability and improvement strategies: Integrating the perspectives of passengers and transit agencies in North America.* Transport Reviews, 2015. **35**(3): p. 292-328.

2. Arhin, S., et al., *Optimization of transit total bus stop time models.* Journal of Traffic and Transportation Engineering (English Edition), 2016. **3**(2): p. 146-153.

3. Pi, X., et al., *Understanding Transit System Performance Using AVL-APC Data: An Analytics Platform with Case Studies for the Pittsburgh Region.* Journal of Public Transportation, 2018. **21**(2): p. 2.

4. Grant, M., *State DOT public transportation performance measures: state of the practice and future needs.* Vol. 675. 2011: Transportation Research Board.

5. Congressional Budget Office, *Public Spending on Transportation and Water Infrastructure, 1956 to 2017.* 2018. p. 1-26.

6. Toole, O.R., *Charting Public Transit's Decline*, in *CATO Institute*. 2018.

7. Bai, C., et al., *Dynamic bus travel time prediction models on road with multiple bus routes.* Computational intelligence and neuroscience, 2015. **2015**: p. 63.

8. El-Geneidy, A.M., J. Horning, and K.J. Krizek, *Analyzing transit service reliability using detailed data from automatic vehicular locator systems.* Journal of Advanced Transportation, 2011. **45**(1): p. 66-79.

9. Murray, A.T. and X. Wu, *Accessibility tradeoffs in public transit planning.* Journal of Geographical Systems, 2003. **5**(1): p. 93-107.

10. Fan, Y., A. Guthrie, and D. Levinson, *Waiting time perceptions at transit stops and stations: Effects of basic amenities, gender, and security.* Transportation Research Part A: Policy and Practice, 2016. **88**: p. 251-264.

11. Strathman, J.G., et al., *Automated bus dispatching, operations control, and service reliability: Baseline analysis.* Transportation Research Record, 1999. **1666**(1): p. 28-36.

12. TransitCenter, *Your Bus Is On Time. What Does That Even Mean?* 2018.

13. Cevallos, F., *Transit Service Reliability: Analyzing Automatic Vehicle Location (AVL) Data for On-Time Performance and Identifying Conditions Leading to Service Degradation.* 2016.

14.     Los Angeles County Metropolitan Transportation Authority, *2016 Metro Transit Service Policies & Standards*. 2015, LA Streetsblog.

15.     Chicago Transit Authority, *Title VI Program Triennial Report 2013-2015*. 2016.

16.     Metro, K.C. *Service Quality Monthly Performance Measures*. 2018; Available from: https://kingcounty.gov/depts/transportation/metro/about/accountability-center/performance/service-quality.aspx.

17.     RTD - Denver. *2016 Performance report-fourth quarter*. 2017.

18.     TRIMET. *Performance Dashboard: March 2019*. 2019; Available from: https://trimet.org/about/dashboard/index.htm#efficiency,%20March%202019.

19.     Metropolitan Council, *2009 Twin Cities Transit System Performance Evaluation*. 2010. p. 1 - 125.

20.     Miami Dade County, *Performance Report for Miami-Dade County Bus Routes Outsourced to Limousines of South Florida, Inc*. 2018.

21.     Port Authority of Allegheny County, *Annual Service Report 2017*. 2018.

22.     Valley Metro, *Transit Performance Report FY 2018*. 2019.

23.     NORTA. *Performance Dashboard*. 2019; Available from: https://www.norta.com/performance.

24.     SEPTA, *Service Standards And Process*. 2016.

25.     Capital Metro, *Metro Performance Dashboard: Reliability.* 2019.

26.     NFTA-Metro, *2017-2018 Annual Performance Report: Key Performance Indicators*. 2018.

27.     SFMTA, *Muni On-Time Performance*. 2019.

28.     Bi-State Development, *Operating & Capital Budget: Fiscal year 2017*. 2018.

29.     City of Charlotte, *CATS On Time Performance*. 2019; Available from: https://charlottenc.gov/cats/bus/Pages/on-time.aspx.

30.     Ride UTA. *Percent of On-Time Performance*. Available from: https://www.rideuta.com/About-UTA/Community-Service-Standards/Percent-of-On-Time-Performance.

31.     WMATA, *Q2 FY2019 Metro Performance Report*. 2019. p. 1-29.

32.     Maryland Transit Administration, *MDOT MTA Performance Improvement*. 2019; Available from: https://www.mta.maryland.gov/performance-improvement.

33.     Fabian, J.J., G.E. Sánchez-Martínez, and J.P. Attanucci, *Improving High-Frequency Transit Performance through Headway-Based Dispatching: Development and Implementation of a Real-Time Decision-Support System on a Multi-Branch Light Rail Line.* Transportation Research Record, 2018. **2672**(8): p. 363-373.

34.   NJ Transit, *On-time Performance*. 2019; Available from: https://www.njtransit.com/var/var_servlet.srv?hdnPageAction=ScoreCardCE2To.

35.   METRO, *Fiscal Year 2019 Monthly Performance Report: Revenue • Expense • Ridership • Performance*. 2019.

36.   MARTA. *Key Performance Indicators*. 2019; Available from: https://www.itsmarta.com/kpihome.aspx.

37.   RTCSNV, *Minutes Regional Transportation Commission Of Southern Nevada* 2018.

38.   MBTA. *Dashboard: How are we measuring reliability?* 2019  07/27/2019]; Available from: https://www.mbtabackontrack.com/performance/index.html#/explanations.

39.   City and County of Honolulu, *Title VI Program Report*. 2015.

40.   MTA. *Bus Performance Dashboard*. 2019  07/27/2019]; Available from: http://busdashboard.mta.info/.

41.   Reed, T. and J. Kidd, *INRIX - Global Traffic Scorecard*. 2019.

42.   TRIMET. *Improving On-Time Performance*. 07/27/2019]; Available from: https://trimet.org/betterbus/#ontime.

43.   Miller, A., *From the Bus Stop to the Fast Lane: How Cities Can Speed up Buses, Improve Ridership*. 2019, Streetsblog Denver.

44.   Baker, R.J., et al., *An overview of transit signal priority*. 2002.

45.   Hemingson, T., *MetroRapid Transit Signal Priority — Using Technology to Improve Service Quality*. 2015.

46.   Schmitt, A., *Richmond Shows How to Boost Small-City Transit*. 2019, Streetsblog USA.

47.   NACTO., *Better Boarding, Better Buses: Streamlining Boarding & Fares*. 2017.

48.   Levy, A. *Here are four achievable ways to build a better bus system*. 2017  07/27/2019]; Available from:   https://ggwash.org/view/65004/here-are-four-achievable-ways-to-build-a-better-bus-system.

49.   Philipsen, K. *Ten Ways to Improve Bus Transit Use and Experience*.  07/27/2019]; Available from: https://www.smartcitiesdive.com/ex/sustainablecitiescollective/ten-ways-improve-bus-transit/1074286/.

50.   Bliss, L. *Why Seattle Is America's Bus-Lovingest Town*. 2018  07/27/2019]; Available from: https://www.citylab.com/transportation/2018/05/seattle-the-city-that-respects-the-power-of-the-bus/559697/.